\documentclass[11pt,showkeys]{revtex4-1}
\usepackage{epsfig,subfigure,amsmath,amssymb}
\begin{document}
\title{Energy and Angular Momentum Storage in a Rotating Magnet}

\author{H. S. Mani}
\affiliation{Chennai Mathematical Institute,\\  Plot H1, SIPCOT IT Park,\\ Padur PO, Siruseri 603103, India}
\author{Praveen Pathak and Vijay A. Singh \email{praveen@hbcse.tifr.res.in}}
 \affiliation{ Homi  Bhabha Centre For Science  Education (TIFR),\\ V.
 N.   Purav Marg,  Mankhurd,\\  Mumbai -  400088, India.   }

\date{\today}

\begin{abstract}
We   consider a  cylindrical metallic  magnet that  is
set into rotation about a horizontal axis by a falling mass. In such a
system the magnetic field will cause a radial current which is
non-solenoidal. This leads to charge accumulation and a partial
attenuation of radial current. The magnetic field acting  on the
radially flowing current  slows down the 
acceleration. Using newtonian dynamics we evaluate the angular
velocity and displacement. We  explicitly show that the
electromagnetic angular momentum must  be taken into consideration in
order to account for the change in angular momentum due to the
external torque on the system. Further the loss in potential energy of
the falling mass can be accounted for only after taking into
consideration the electrostatic  energy and the Joule loss.  We
suggest that this example will be of pedagogical value to intermediate
physics students.  A version of this is scheduled to appear in American
Journal of Physics 2011.

 \end{abstract}

\keywords{Rotating magnet, Charge accumulation}

\maketitle
\section{Introduction}

Introductory and  intermediate treatments of  electromagnetism usually
shy away from a  formal discussion of electromagnetic angular momentum
and energy.\cite{hall92,serw94,gian00} Discussions if any are kept at
a qualitative level. Specific examples involving forces and torques as
well as  the associated work done  and the change  in angular momentum
are conspicuous by their absence. In a series of articles Gauthier has
attempted to redress this lacuna.\cite{gaut82,gaut02}

In one of his articles Gauthier considers a metallic cylindrical shell
free to rotate about the horizontal axis.\cite{gaut02} A string wound
around the  shell with a  mass attached to  its free end causes  it to
rotate with acceleration.  A charge is smeared on  the shell making it
analogous  to  a solenoid  carrying   an  increasing time  dependent
current and associated magnetic flux.  One can show from Faraday's law
that a  retarding torque  will act  on the shell.   This torque  is of
purely electromagnetic origin.  Gauthier  goes on to demonstrate in a
pedagogically satisfying fashion that the system's magnetic energy and
electromagnetic momentum are stored at the expense of their mechanical
counterparts.

In the present article we consider a cylindrical magnet free to rotate
about the  horizontal axis.  We treat the magnetic 
field of the magnet to be  constant and pointing along the axis.  A
string wound  around the shell  with a 
mass   attached  to   its  free   end   causes  it   to  rotate   with
acceleration. We assume that the  magnet is a good conductor and hence
a  radial current  will be  set up. The magnetic field will act on the
radial current giving rise
to a counter torque which will oppose the torque due to
gravity.\cite{cylinder_inB}  The  
problem  is  complementary to 
Gauthier's. Instead  of Faraday's  law we need  to invoke  the Lorentz
force, Gauss's law and the equation of continuity. But the perspective
is the same as Gauthier's:  an understanding of the problem involves a
detailed  consideration  of  the  electromagnetic energy  and  angular
momentum. We  thus present another pedagogical example  for an intermediate
electromagnetic course.

\section{Dynamics of a rotating magnet}\label{sec:dyna}
\begin{figure}[h]
 \begin{center}
\subfigure[] {\epsfig{file=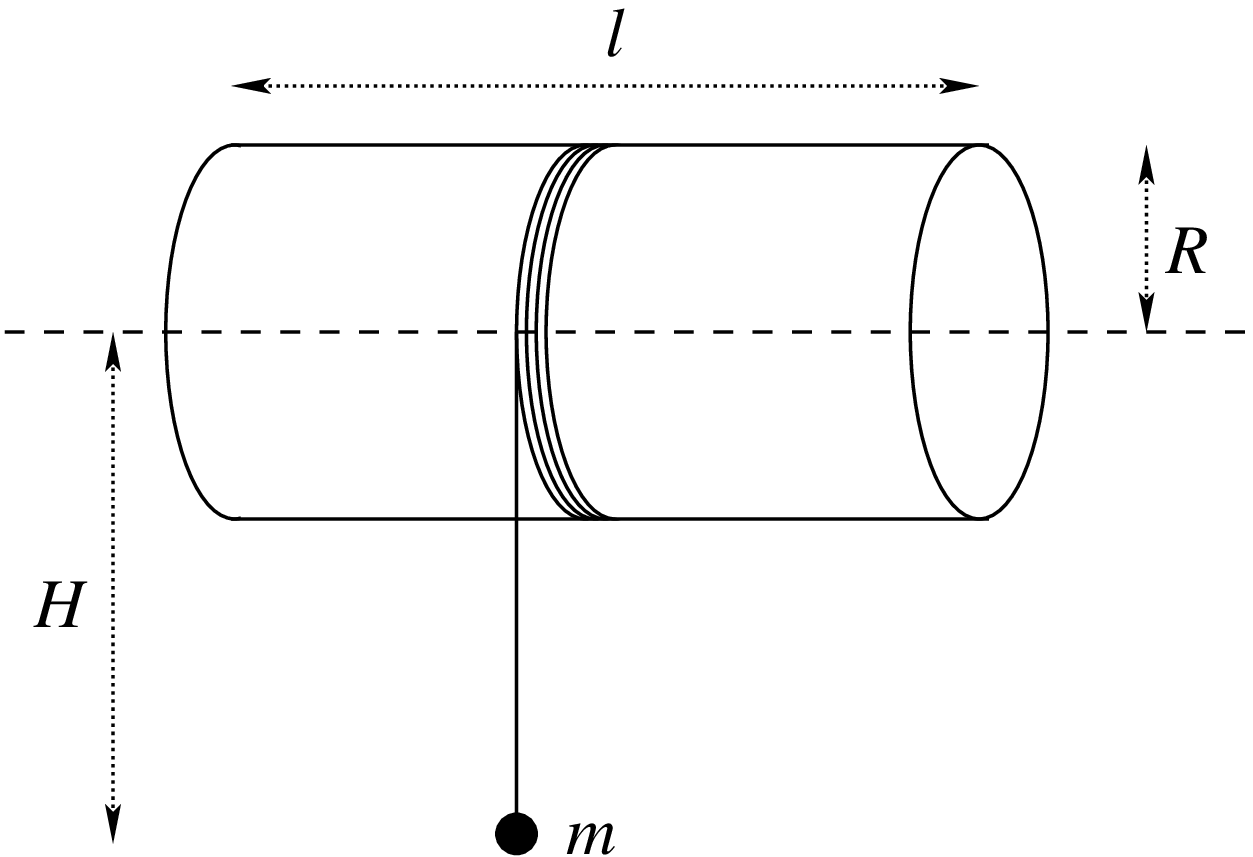,height=3.5cm}\label{fig:cylinder}} 
\subfigure[] {\epsfig{file=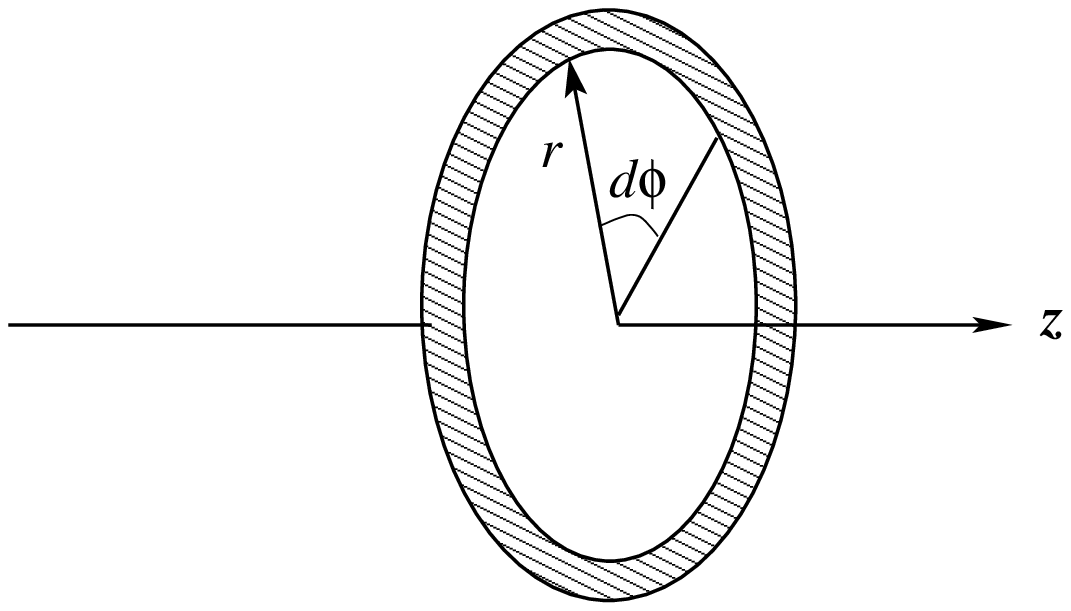,height=3cm}\label{fig:cross_sec}}
\caption{(a) Setup. (b) The cylindrical coordinate system. See text
  for details.}
\end{center}
\end{figure}

Consider a cylinder with moment  of inertia $I$, radius $R$ and length
$l$. Let mass  $m$ be hung by  a light thread wound on  the surface of
the   cylinder.    We   adopt   the  cylindrical   coordinate   system
\{$r,\phi,z$\} and unit vectors \{$\hat  r, \hat \phi, \hat k$\} with
$\hat  k$ being  the axis  around which  cylinder is  free  to rotate.
Figure   1(a)     illustrates     the    setup     and
Fig.  1(b) the  cylindrical coordinate  system  we have
adopted.   When  mass  $m$  is  released, the  cylinder  rotates  with
instantaneous  angular  velocity $\omega'(t)  \hat  k$.  We assume no
slip condition $v=r\omega'$. In that  case
 Newton's second law for rotational dynamics can be used to show that   angular velocity of
the  cylinder   is \cite{gaut02}
 \begin{equation}  \omega'=  \frac{\beta}{\tau_1^2}
  t,  \label{eq:omegacyl}\end{equation}  where  we  have  introduced  a
dimensionless constant  \begin{equation}
 \beta  = \dfrac{mR^2}{I+mR^2} \label{eq:omega'},\end{equation}
 and  a time
constant \begin{equation} \dfrac{1}{\tau_1^2}  =
  \dfrac{g}{R}. \label{eq:tau1} \end{equation}  

We  now replace  the  cylinder by a metallic cylindrical magnet in the setup. Let  the
conductivity of the cylinder be $\sigma$
and the  magnetic field be uniform and  $B\hat k$.\cite{cylinder_inB}  This is
similar  to the  setup described  by Gauthier \cite{gaut02} except
that we have a 
solid cylindrical magnet instead of a 
non-magnetic charged cylindrical  shell. As the magnet
rotates due to the gravitational torque of the falling mass, the free
charge in the rotating magnet experiences a radially outward Lorentz
force. The magnetic field will act on this radial current and will
produce a counter torque ($\tau_{em}$) which will oppose the torque due to
gravity. The subscript indicates  that the torque is of purely  electromagnetic
origin. We can  write the equation of motion for 
the system  as
\begin{equation} 
m g R - \tau_{em}(t) = (I+m R^2) \frac{d\omega}{dt}. \label{eq:eqmotion1}
\end{equation}

In what follows we will  determine $\tau_{em}$ and solve the equation
of motion.

Consider a free charge $q$ at a distance $r\  (r<R)$  from the axis of
the cylinder rotating with angular velocity $\omega$. It is subjected
to a radial Lorentz force $q\omega r B \hat{r}$.  Recall that in 
literature the effect of this force is  discussed in 
terms of motional emf. \cite{griffith,galili06}
 From Ohm's law this motion  would lead to a current density
 \(\vec{J}_m = \sigma \omega r B \hat r\). 
It is important
to note that this current density is non-solenoidal 
\[ \vec\nabla.\vec J_m = \frac{1}{r}\frac{d}{dr}(rJ_m) \neq 0\]  and hence 
\textit{charge must  accumulate}.
 This accumulating  charge  will
give  rise to  a 
electric  field ($\vec E_a$).   Let
$\rho(\vec r, t)$ be  the charge distribution due  to accumulation. Associated
with this $\rho(\vec r, t)$ is an  electric field which can  be determined
by   Gauss's law  with the standard pill box 
construction in the magnet \cite{hall92}
\[ E_a = \frac{1}{\epsilon_0 r}\int^r_0{\rho(r',t)r'
    dr'} \ \ \ \ (r<R), \]  where we assume that  the length of  magnet
$l\gg R$. 
Once again we use  Ohm's law to obtain  $\vec J_a=\sigma  E_a \hat r$. The
resulting current density across
the magnet  is then 
 \begin{equation} \vec J 
 = \sigma \left(\omega r B + \frac{1}{\epsilon_0
  r}\int^r_0{\rho(r',t)r' dr'} \right)\hat r,  
  \label{eq:J}\end{equation} 
It consists of two terms: the first is due to the motional emf and second
is due to charge accumulation. Next examine  the divergence of current density  \[ \vec\nabla.\vec J =  \frac{1}{r}\frac{d}{dr}(rJ) = \sigma\left(2\omega B +
\frac{1}{\epsilon_0}\rho(r,t)\right)   \]

From equation of continuity
\begin{equation*} 
 \vec\nabla.\vec    J    +   \frac{\partial\rho}{\partial    t}   =
 0 \end{equation*} 
we obtain 
\begin{equation}  \frac{\partial \rho}{\partial  t} +
 \frac{\sigma\rho}{\epsilon_0}                                        =
 -2B\sigma\omega(t).  \label{eq:contu}  \end{equation} 
  
We assume  that initially the cylinder  is neutral (\(\rho(\vec r,t=0)
=  0\)). Using this, 
Eq. (\ref{eq:contu}) can be solved to yield
\begin{equation} \rho(r,t) = -2B\sigma e^{-t/ \tau_2}
\int^t_0{\omega(t')e^{t'/\tau_2}dt'}, \label{eq:rho} \end{equation}  where
$\tau_2=\epsilon_0/\sigma$.  The reader may have encountered $\tau_2$
earlier. When a lump of charge 
 is  deposited on a conductor, it  spreads out in a very short time
 and with time constant $\tau_2$.  Note  that
$\rho$ is independent of $r$. In order to maintain electrical
neutrality there will be an accumulation of charges at the
surface with the sign opposite to $\rho$. Hence
\begin{equation} 
E_a=0\ \ \ \ \ \ \ \ (r>R). \label{eq:Ea}
\end{equation} 
Consider an annular element of radius $r <R $ in the cylinder (see
Fig. \ref{fig:cross_sec}). The force ($d\vec F$) 
  on the annular element can be written as
\begin{eqnarray*}
d\vec F & =& 2\pi  l  r   J d\vec{r}  \times \vec B\\
        & =& 2\pi  l r J B (-\hat \phi) dr
\end{eqnarray*}
The associated torque ($d\vec\tau_{em}$) on the ring will be 
\begin{equation*} 
d\vec\tau_{em}(t) =  \vec r \times d\vec F = 2\pi r^2 l J B (-\hat k) dr
\end{equation*}
The subscript indicates that torque is of electromagnetic origin. The net
torque is obtained by straight forward integration  
\begin{equation*} 
d\vec\tau_{em}(t) =  \vec r \times d\vec F = 2\pi r^2 l J B (-\hat k) dr.
\end{equation*}
Using the expressions for $J$ (Eq.  (\ref{eq:J}))   and
 $\rho$ (Eq. (\ref{eq:rho}))  one may obtain the net torque after
straight forward  integration
\begin{equation} 
\tau_{em}(t)= \frac{\pi l \sigma B^2 R^4}{2} \left(\omega(t) -
\frac{1}{\tau_2} e^{-t/\tau_2} \int^t_0{\omega(t')e^{t'/\tau_2}
  dt'}\right). \label{eq:tau}
\end{equation} 
We pause to note that since  the charge density is independent of $r$,
the electric field  from Gauss's law will have  a linear dependence on
$r$. Ohm's law implies that the current density would also be linearly
dependent on  $r$.  It now  remains to determine the  time dependence.
Differentiating  above Eq.  (\ref{eq:eqmotion1}) with  respect  to $t$
gives
\begin{equation} 
\frac{d^2\omega}{dt^2}  = \frac{1}{(I+m
  R^2)}\frac{d}{dt}\tau_{em} \label{eq:eqmotion2}   
\end{equation} 
Using  Eq. (\ref{eq:tau})
\begin{equation} 
\frac{d}{dt}\tau_{em}= \frac{\pi l \sigma B^2 R^4}{2} \left(\frac{d\omega}{dt} -
\frac{1}{\tau_2} \left( -\frac{1}{\tau_2}e^{-t/\tau_2} \int^t_0{\omega(t')e^{t'/\tau_2}
  dt'} + \omega(t) \right)\right) \label{eq:dtau}
\end{equation} 
The  integral term on the r.h.s of   Eq. (\ref{eq:dtau})  can  be eliminated  using
Eqs.  (\ref{eq:tau})  and  (\ref{eq:eqmotion1}).  Hence we may   rewrite  Eq.
(\ref{eq:eqmotion2}) as
\begin{equation} 
\frac{d^2\omega}{d t^2} + \left(\frac{\pi l \sigma B^2 R^4}{2(I+ mR^2)} +
\frac{1}{\tau_2} \right) \frac{d\omega}{dt} =
\frac{mgR}{\tau_2(I+mR^2)}.  \label{eq:omegamag1}
\end{equation}
We define a  time constant \begin{equation}  
\frac{1}{\tau_3} = \frac{\pi l  B^2 \sigma
  R^4}{2(I+mR^2)},\label{eq:tau3}\end{equation} 
 and an effective time constant
\begin{equation*}  \frac{1}{\tau_e} =
\frac{1}{\tau_2} + \frac{1}{\tau_3}.
\end{equation*} 
One can solve Eq. (\ref{eq:omegamag1}) using
initial  condition \(\omega(t=0) = 0\) and \(d\omega/dt =
\beta/\tau_1^2 \) to obtain  
\begin{eqnarray} 
\omega(t) = \frac{\beta}{\tau_1^2} t - \frac{\beta\tau_e}{\tau_1^2
  \tau_3} f(t)\label{eq:omegamag2} \\
\mbox{where} \ \ \ f(t) = t-\tau_e\left(1-e^{-t/\tau_e}\right) \label{eq:f}
\end{eqnarray} 
Note that the first term in Eq. (\ref{eq:omegamag2}) is same as
Eq. (\ref{eq:omega'}).   Also  \[f(t)>0\] This indicates that the angular
velocity in case of magnet is never greater than in the case of
non-magnetic solid cylinder.   We consider two limits of the
Eq. (\ref{eq:omegamag2}) 
\begin{eqnarray}
\omega(t\ll \tau_e )& \approx & \frac{\beta }{\tau_1^2} t -
\frac{\beta}{2 \tau_1^2
  \tau_3}t^2 \\
\omega(t \gg \tau_e)& \approx & \frac{\beta \tau_e}{\tau_1^2\tau_2}
t \\& <& \frac{\beta}{\tau_1^2} t = \omega'
\end{eqnarray}
This shows that  the effect of the metallic magnet  is to decrease the
angular   velocity  from  the   non-magnetic  value   ($\omega'$).   A
straightforward integration also yields the angular distance traversed
\begin{eqnarray} 
\phi(t) = \frac{\beta \tau_e^2}{\tau_1^2\tau_3}f(t) + \frac{\beta
  \tau_e}{2\tau_1^2\tau_2} t^2 \label{eq:phi}
\end{eqnarray}
 Once again we note the limiting
cases 
\begin{eqnarray*}
\phi(t\ll \tau_e)& =& \frac{\beta}{2\tau_1^2}t^2\\
\phi(t\gg \tau_e) &=& \frac{\beta \tau_e^2}{\tau_1^2\tau_3}t+\frac{\beta \tau_e}{2\tau_1^2\tau_2} t^2\\
&<& \frac{\beta}{2\tau_1^2} t^2
\end{eqnarray*}
which  indicates that  the  magnetic field  slows  down the
rotation. Using Eq.  (\ref{eq:omegamag2}) in Eq.  (\ref{eq:rho}) and integrating
we obtain charge density
\begin{equation} 
\rho(t)= \frac{-2B\sigma\beta\tau_e}{\tau_1^2}f(t).
\end{equation} 
We  remind the  reader that  $f(t) >  0$ and  hence charge  density is
negative.\cite{chargedensity} Also, as stated earlier, a  compensating
positive charge collects 
on the surface so that the magnet is on the whole  neutral.   We  have
already pointed  out  that $\rho$  is independent of  $r$ and  therefore the
electric field  is
\begin{eqnarray}
\vec E_a = \frac{\rho(t)r}{2\epsilon_0} \hat r &=& -\frac{B\beta\tau_e
r}{\tau_1^2\tau_2}f(t) \hat r\ \ \ \ \ (r<R), \label{eq:finalEa}\\
&=& 0 \ \ \ \ \ \ \ \ \ \ \ \  \ \ \ \ \ \ \ \ \ \ (r>R).
\end{eqnarray}

We then calculate  the total current density 
\begin{equation} 
\vec{J} = \frac{\sigma r B\beta}{\tau_1^2} \left(t-f(t)\right)\hat{r}. \label{eq:finalJ}
\end{equation} 
while the net  electromagnetic torque on the system is 
\begin{equation} 
\tau_{em} = \frac{\pi \epsilon_0 l B^2 R^4 \beta}{2 \tau_1^2 \tau_2}(t-f(t)). \label{eq:net_torque}
\end{equation} 
We pause to comment briefly on the time dependence of above
quantities. Note that \mbox{\(t-f(t) = \tau_e(1-e^{-t/\tau_e})\)}. As
physically expected both the current density and electromagnetic
torque are initially zero and they both saturate to a constant value at
large times.

\section{verification of angular momentum}
The electromagnetic angular momentum ($L_{em}$) is defined in terms of
electric and magnetic  field.\cite{griffith} In our case we have  
\begin{eqnarray}  \vec L_{em} = \int^R_0{\epsilon_0}\vec{r'} \times(\vec E_a\times \vec
B)\,2\pi r'l dr'\label{eq:angudef} \end{eqnarray}
 
Using the expression for electric field due to the charge accumulation (Eq. (\ref{eq:finalEa})) we have,
\begin{eqnarray}
\vec L_{em} & = &  - \frac{\epsilon_0 B^2    \beta \tau_e  2
  \pi l}{{\tau_1}^2\tau_2}f(t) \int^R_0 r' \hat r \times (r' \hat r \times
\hat k) r' dr' \nonumber\\
  & = & \frac{\epsilon_0 B^2    \beta \tau_e  2
  \pi l}{{\tau_1}^2\tau_2}f(t) \int^R_0 r'^3 dr' \hat k \nonumber\\
 & = & \frac{\epsilon_0 B^2    \beta \tau_e\pi l  R^4 
  }{2{\tau_1}^2\tau_2}f(t)\hat k \label{eq:emangu}
\end{eqnarray}

Since $f(t)$ is always positive it means that some angular momentum is
transferred  to   the  electromagnetic  field   and  consequently  the
mechanical  angular  momentum  is   diminished.  One  can  check  that
differentiating  the  above  equation  for $L_{em}$,  we  will  obtain
precisely  Eq. (\ref{eq:net_torque})  for electromagnetic  torque. The
mechanical angular momentum of the system is
\begin{equation} 
\vec L_m =(I+mR^2)\omega \hat k =(I+mR^2) \left(\frac{\beta}{\tau_1^2} t - \frac{\beta\tau_e}{\tau_1^2
  \tau_3} f(t)\right) \hat k.
\label{eq:mechangu}\end{equation}
Adding Eqs. (\ref{eq:emangu}) and (\ref{eq:mechangu}), we see 
that the  terms associated with  $f(t)$ cancel and lead to the expression 
\begin{equation} \vec L_{em}+ \vec L_m = mgR t \hat k.\end{equation} 
as it  should, being  the change in  angular momentum due  to the external
torque on the system.

\section{Energy Balance}
As the hanging mass drops a distance $H$, the loss in potential
energy is
\begin{equation} 
U = mgH = mgR\phi(t),
\end{equation} 
where $\phi(t)$ is given by Eq. (\ref{eq:phi}). At first sight it may
seem that   this 
potential  energy is   distributed into two channels : (i)
gain in  kinetic energy of system (mass+cylinder)  and (ii) electrical
energy stored in the system. Let us evaluate the contributions of
these two channels.  

The  total kinetic energy after the hanging mass has dropped a distance
$H=R\phi(t)$ is 
\begin{eqnarray}
 K &=& \frac{1}{2}I\omega^2 + \frac{1}{2}mv^2 \nonumber\\
   &=& \frac{I+mR^2}{2}\omega^2. \label{eq:mechenergy}
\end{eqnarray}
where $v$ is the linear speed  of the  falling mass and $\omega$ is
given by Eq. (\ref{eq:omegamag2}). On the other hand the stored  electrical energy ($U_e$) is 
 \begin{eqnarray} 
U_e &=& \frac{\epsilon_0}{2} \int^R_0{E_a^2\,2\pi l
  r'\,dr'} \label{eq:uedefi} \\
& =& \frac{\epsilon_0 \pi l B^2 \beta^2\tau_e^2R^4
}{4\tau_1^4\tau_2^2}f^2(t). \label{eq:energyelec}  
\end{eqnarray}  
Note that the integration is up to $R$ only (See
Eq. (\ref{eq:Ea})).  However adding Eqs. (\ref{eq:mechenergy}) and
(\ref{eq:energyelec}) yields  
\begin{equation}
U-(K+U_e) = (I+mR^2)\frac{\beta^2\tau_e}{2 \tau_1^4
  \tau_3} \left[ 2 \tau_e f(t) - \left(f(t)- t\right)^2\right]. \label{eq:diffE}
\end{equation}
The right hand side is evidently not zero. We need to evaluate the
Joule loss. This is given by  
\begin{eqnarray} 
 U_{J}&  =& \frac{1}{\sigma}\int^t_0{\int^R_0 {J^2\,2\pi  l r'  \, dr'
     dt'}}, \label{eq:Uj}  \end{eqnarray} where the current density is
given by  Eq.
(\ref{eq:finalJ}).  The heat loss in a conductor is due
to the collision of the charge carriers with the surrounding
atoms. The collisions 
result   in the  dissipation of   kinetic energy of
the charge carriers in the
form of heat.  This motion of charge carriers  arises from two
mutually opposing processes: (i) 
 the electric field $E_a$ given by Eq. (\ref{eq:finalEa}) and (ii)
 the Lorentz force. Note that  in the evaluation of angular momentum  (Eq. (\ref{eq:angudef}))  we considered
only  electric  field ($  E_a $).  The point to appreciate is that in
the case of energy conservation we need to also consider the Lorentz
force since it results in the motion of charge with consequent  loss
in kinetic energy due to collisions.  Substituting the current density from Eq.
(\ref{eq:finalJ}) and using Eq. (\ref{eq:f}), we obtain
\begin{eqnarray}
 U_J & = & \frac{\sigma B^2 \beta^2  \pi l \tau_e^2 R^4}{ 2 \tau_1^4}
 \int^t_0  \tau^2_e(1-e^{-t'/\tau_e})^2  dt'.  
\end{eqnarray}  $U_J$ can be evaluated
by lengthy  and straightforward  integration (See Appendix for
detailed derivation of $U_J$). It yields  precisely the
r.h.s. of  Eq. (\ref{eq:diffE}). Thus \[ mgH = K + U_e +U_J. \] 
\section{Discussion}
Let  us recapitulate the  physical processes which are involved. The magnet
rotates due to the gravitational torque of the falling mass. The free
charge in the rotating  magnet experiences a radially outward force. 
The  resulting radial  current  is non-solenoidal.   This implies  that
charge will accumulate.  The  electric field due to the charge
accumulation will retard the 
radial motion  of free charge. The resulting   diminished radial current
will nevertheless  give rise to  a counter-torque which is of
patently 
electromagnetic origin. Newton's second law can be applied to derive
the angular velocity and angular acceleration. Next, we demonstrate by
detailed bookkeeping that energy and angular momentum are properly accounted
for. 

 One  may get  a  perspective on  the  problem by  obtaining order  of
 magnitude  estimates of  time  constants involved.  Let  us take  the
 radius of the  magnet $R$= 0.1 m,  length $l$ = 1 m,  density same as
 that of  iron (7800 kg$\cdot$m$^{-3}$), conductivity  $\sigma = 10^7$
 S$\cdot$m$^{-1}$ and  magnetic field $B$ to  be 1 T.  Let the falling
 mass      be     10      kg.     This      yields     \(\tau_1\approx
 10^{-1}\,\mbox{s},\ \tau_2\approx 10^{-18}\,\mbox{s},\ \tau_3 \approx
 10^{-3}\,\mbox{s}\) and \(  \tau_e \approx 10^{-18}\,\mbox{s}\). Note
 that  $\tau_e \cong  \tau_2$. When  one uses  these numbers  then one
 finds  that current  density is  very small  and of  order 10$^{-12}$
 A$\cdot$m$^{-2}$.   Thus  the  motion  of  the charges  will  have  a
 negligible effect on the  magnetic field.  The observable effects are
 possible  if the  magnetic field  is  unusually large  as happens  in
 magnetars with magnetic  field of gigateslas.\cite{macd00} However it
 is  important to  realise that  howsoever  small the  effects due  to
 magnetic field may be they  are crucial in accounting for the angular
 momentum and energy balance.
\begin{acknowledgments}
\noindent
This  work  was  supported   by  the  Physics  Olympiad  and  National
Initiative  on Undergraduate  Sciences (NIUS)  undertaken by  the Homi
Bhabha Centre for Science Education (HBCSE-TIFR), Mumbai, India.
\end{acknowledgments}
\appendix*
\section{}
\label{sec:appendix}
We substitute  the expression for current  density from
Eq. (\ref{eq:finalJ}) into Eq. (\ref{eq:Uj}) and get
\begin{eqnarray}
U_J& = & \frac{\sigma B^2 \beta^2 2 \pi l \tau_e^2}{\tau_1^4}\int^R_0
 r'^3 dr'\int^t_0 \left( t'- f(t') \right)^2 dt' 
\end{eqnarray}
The integrand associated with the  $t$ variable can be 
simplified with the help of Eq. (\ref{eq:f}) for $f(t)$ and this yields
\begin{eqnarray}
 U_J & = & \frac{\sigma B^2 \beta^2  \pi l \tau_e^2 R^4}{ 2 \tau_1^4}
 \int^t_0  \tau^2_e(1-e^{-t'/\tau_e})^2  dt' \nonumber 
\end{eqnarray} 
Expanding and integrating the squared term and using the time constant $\tau_3$ (Eq. (\ref{eq:tau3}))  
\begin{eqnarray}
 U_J& =& (I+mR^2)\frac{\beta^2\tau_e^2}{ \tau_1^4 \tau_3}
 \left( t+\frac{\tau_e}{2} (1-e^{-2t/\tau_e}) -2\tau_e
 (1-e^{-t/\tau_e})\right) \nonumber
\end{eqnarray}
Using Eq. (\ref{eq:f}) once again
\begin{eqnarray}
U_J& =& (I+mR^2)\frac{\beta^2\tau_e^2}{ \tau_1^4 \tau_3}
 \left( t+\frac{t-f(t)}{2}\left(2-\frac{t-f(t)}{\tau_e} \right) -2
 (t-f(t))\right) \nonumber \\ 
& =& (I+mR^2)\frac{\beta^2\tau_e^2}{ \tau_1^4 \tau_3}\left( f(t) -
 \frac{\left( t- f(t)\right)^2}{2\tau_e}\right) \label{eq:finalUj}
\end{eqnarray}

Equation (\ref{eq:finalUj}) is  precisely the r.h.s. of
Eq. (\ref{eq:diffE}).


\begin{thebibliography}{10}
\bibitem{cylinder_inB}Instead of a magnet we can also consider a long
  metallic cylinder placed in a uniform magnetic field directed along the axis
  of the cylinder.
\bibitem{chargedensity} Note that the increase of charge with time
  cannot happen indefinitely. We have assumed in our model that
  conductivity is constant.
\end{thebibliography}
\end{document}